\documentclass[12pt]{article}
\usepackage{amsmath,amssymb,graphicx,mathbbol}

\def\appendix#1{
  \addtocounter{section}{1}
 \setcounter{equation}{0}
  \renewcommand{\thesection}{\Alph{section}}
 \section*{Appendix \thesection\protect\indent \parbox[t]{11.715cm} {#1}}
  \addcontentsline{toc}{section}{Appendix \thesection\ \ \ #1}
  }

\def\dm{\Delta_{\rm max}}
\def\4N{$\mathcal{N}=4$}

\begin{document}

\thispagestyle{empty}
\begin{flushright}%\footnotesize\tt
%hep-th/0512079\\
ITEP-TH-105/05\\
UUITP-26/05
\end{flushright}
\vspace{.5cm} \setcounter{footnote}{0}
\begin{center}
{\Large{\bf Antiferromagnetic Operators in N=4 Supersymmetric
Yang-Mills Theory
\par}
   }\vspace{1cm}
{\large\rm K. Zarembo\footnote{Also at ITEP, Moscow, 117259
Bol. Cheremushkinskaya 25, Moscow, Russia}}\\[5mm]
{\it Institutionen f\"or Teoretisk Fysik,
Uppsala Universitet\\
Box 803, SE-751 08 Uppsala,  Sweden}\\[2mm]

%\vskip 1cm

{\tt\noindent  konstantin.zarembo@teorfys.uu.se}

\vskip 2.5cm

%\newpage
{\sc Abstract}\\[2mm]
\end{center}
\noindent The spectrum of operators in the $su(2)$ sector of
$\mathcal{N}=4$ SYM is bounded because the number of operators is
finite. According to the AdS/CFT correspondence, the string spectrum
in this sector should be also bounded. In this paper the upper bound
on the scaling dimension is calculated in the limit of the large
R-charge using Bethe ansatz.
\newpage
\setcounter{page}{1}
\renewcommand{\thefootnote}{\arabic{footnote}}
\setcounter{footnote}{0}

The AdS/CFT correspondence \cite{Maldacena:1998re} predicts that the
spectrum of local gauge-invariant operators in $\mathcal{N}=4$ SYM
theory and the string spectrum in $AdS_5\times S^5$ are identical
\cite{Gubser:1998bc,Witten:1998qj}. One puzzling moment in this
identification is that the set of operators superficially looks more
discrete than the set of string states. This point is best
illustrated by considering a particular set of operators
\begin{equation}\label{oops}
 \mathcal{O}={\mathop{\mathrm{tr}}\nolimits}(Z^{J_1}W^{J_2}
 +{\rm permutations}),
\end{equation}
where $Z$ and $W$ are two complex scalar fields from $\mathcal{N}=4$
supermultiplet. This set of operators is closed under
renormalization \cite{Beisert:2003jj}, but mixing of the operators
among themselves is non-trivial and is best described by mapping to
a quantum spin chain of length $L=J_1+J_2$ \cite{Minahan:2002ve}.
Each occurrence of  $Z$ in an operator represents spin up and an
occurrence of $W$ represents spin down. Cyclically symmetric
distributions of spins on a one-dimensional lattice of length $L$
are then in the one-to-one correspondence with all possible
orderings of the fields $Z$ and $W$ under the trace. The planar
dilatation operator, whose eigenvalues are large-$N$ scaling
dimensions of operators (\ref{oops}), can be identified with the
spin-chain Hamiltonian
\cite{Minahan:2002ve,Beisert:2003tq,Serban:2004jf,Beisert:2004hm}.
The ferromagnetic ground state of the spin chain corresponds to the
chiral primary operator ${\mathop{\mathrm{tr}}\nolimits}Z^L$ with
zero anomalous dimension. The excited states (magnons), described by
a collection of spin flips $Z\rightarrow W$ that propagate along the
lattice with momentum $p=2\pi n/L$, correspond to operators with
parametrically small anomalous dimensions (BMN operators
\cite{Berenstein:2002jq}). The contribution of a single magnon to
the anomalous dimension is determined by the dispersion relation
$\varepsilon =(\lambda /2\pi ^2)\sin^2(p/2)+O(\lambda ^2)$, where
$\lambda =g_{YM}^2N$ is the 't~Hooft coupling of the SYM.

It is possible to identify the dual of the $su(2)$ sector
(\ref{oops}) in the classical string theory
\cite{Frolov:2003qc,Kruczenski:2003gt,Kruczenski:2004kw,Kazakov:2004qf},
although not without subtleties \cite{Minahan:2005jq}. The magnons
correspond to transverse fluctuations of the point-like string
orbiting around a big circle of $S^5$ \cite{Gubser:2002tv}. The
dispersion relation for the string modes is $\varepsilon
=\sqrt{1+\lambda p^2/4\pi ^2}-1$. At weak coupling and at small
momenta $p\ll 1$ the spin-chain and the string dispersion relations
match. If we believe in the AdS/CFT correspondence, quantum
corrections on the string side and higher-loop effects in the SYM
should eliminate the difference completely and should produce a
common exact dispersion relation, e.g.
\cite{Beisert:2004hm,Arutyunov:2004vx,Beisert:2005tm}
\begin{equation}\label{1}
 \varepsilon =\sqrt{1+\frac{\lambda }{\pi ^2}\,\sin^2\frac{p}{2}}-1,
\end{equation}
which reduces to the magnon energy at $\lambda \ll 1$ and to the
energy of a classical string mode at $p\ll 1$. There is a small
subtlety here, however. The momentum of a magnon is confined to a
single Brillouin zone : $0\leq p<2\pi $ which together with the
momentum quantization makes the total number of states finite. After
all, there is a finite number  of operators (\ref{oops}) with $L$
fixed\footnote{An asymptotic upper bound on the number of operators
is $2^L$. One can make a better estimate with the help of the Polya
theory \cite{Sundborg:1999ue,Aharony:2003sx}.}. On the contrary, the
momentum of the classical string oscillations is unbounded. Of
course, one can trust the semiclassical approximation only for
sufficiently low world-sheet momenta\footnote{All explicit
calculations on the string side have so far been done only for such
low-momentum states and in fact have been insensitive to the
difference between $p^2/4$ and $\sin^2(p/2)$ in the dispersion
relation \cite{Arutyunov:2005hd} (field-theory calculations,
however, can be pushed beyond the leading order in $p^2$
\cite{Minahan:2005qj,Gromov:2005gp}). I am grateful to S.~Frolov for
this remark.} $p\ll 1$. We currently do not know what happens when
the string is quantized. The quantization can somehow hide states
with the large momentum.

Since the number of operators is finite, their scaling dimensions
are bounded above and can vary in a finite interval
\begin{equation}\label{}
 L\leq \Delta \leq \Delta _{\rm max}.
\end{equation}
As long as the $su(2)$ sector can be identified in the quantum
string theory, the total number of states in it must be also finite
and energies of the string states must satisfy the same bound.

The aim of this note is to calculate $\dm$ in the thermodynamic
limit of $L\rightarrow \infty $. The state with the largest possible
energy is the antiferromagnetic (AF) vacuum of the spin chain. The
numbers of up and down spins in the AF vacuum are equal:
$J_1=J_2=L/2$. One motivation to study this state is that classical
string solutions with the same quantum numbers have been recently
constructed \cite{Park:2005kt,TT}. Another motivation comes from the
study
 of the
spin chain that describes one-loop anomalous dimensions in large-$N$
QCD
\cite{Bukhvostov:1985rn,Braun:1998id,Belitsky:1999bf,Ferretti:2004ba,Belitsky:2004cz,Beisert:2004fv}.
The true ground state there is AF
\cite{Ferretti:2004ba,Beisert:2004fv}. There is much less control
over higher-loop corrections in QCD, not to say over the dual string
theory. It would be thus interesting to study the AF state in the
SYM setting where loop corrections and the string dual are much
better understood.

The scaling dimensions of the operators (\ref{oops}) are eigenvalues
of the dilatation operator
\begin{eqnarray}\label{ham}
 D&=&\sum_{l=1}^{L}\left[
 1
 +\frac{\lambda }{16\pi ^2}\left(1-\boldsymbol{\sigma }_l\cdot\boldsymbol{\sigma }_{l+1}\right)
 -\left(\frac{\lambda }{16\pi ^2}\right)^2\left(
 3-4\boldsymbol{\sigma }_l\cdot\boldsymbol{\sigma }_{l+1}
 +\boldsymbol{\sigma }_l\cdot\boldsymbol{\sigma }_{l+2}\right)
 \right.\nonumber \\
 &&\left.
 +\left(\frac{\lambda }{16\pi ^2}\right)^3\left(
 20-29\boldsymbol{\sigma }_l\cdot\boldsymbol{\sigma }_{l+1}
 +10\boldsymbol{\sigma }_l\cdot\boldsymbol{\sigma }_{l+2}
 -\boldsymbol{\sigma }_l\cdot\boldsymbol{\sigma }_{l+3}
 \right.\right.\nonumber \\ && \left.\left.
 -\boldsymbol{\sigma }_l\cdot\boldsymbol{\sigma}_{l+2}
  \,\,\boldsymbol{\sigma
 }_{l+1}\cdot\boldsymbol{\sigma }_{l+3}
 +\boldsymbol{\sigma }_l\cdot\boldsymbol{\sigma }_{l+3}\,\,\boldsymbol{\sigma
 }_{l+1}\cdot\boldsymbol{\sigma }_{l+2}
 \right)
 \vphantom{\left(\frac{\lambda }{16\pi ^2}\right)^3}
 +O(\lambda ^4)\right]
\end{eqnarray}
At one loop this is the Hamiltonian of the Heisenberg spin chain
\cite{Minahan:2002ve}, which is an integrable model. It is truly
remarkable that the condition of integrability and the requirement
of the BMN scaling imposed on the dispersion relation uniquely fix
the dilatation operator up to $O(\lambda^L )$
\cite{Beisert:2003tq,Beisert:2004hm}. An alternative algebraic
derivation of \cite{Beisert:2003ys,Zwiebel:2005er,Beisert:2005tm}
and explicit three-loop calculations of anomalous dimensions
\cite{Kotikov:2004er,Eden:2004ua} confirm the validity of these
assumptions\footnote{Strictly speaking, the BMN scaling has been
tested only up to three loops. In the plane-wave matrix theory,
which is described by a very similar Hamiltonian, the BMN scaling is
violated at $O(\lambda ^4)$ \cite{Fischbacher:2004iu}.  However, the
plane-wave matrix theory is not exactly integrable
\cite{Beisert:2005wv}.}.

 Integrability of the dilatation operator allows one
to calculate its spectrum with the help of the Bethe ansatz. By
extending the Bethe-ansatz solution of the Heisenberg model
\cite{Bethe:1931hc,Faddeev:1996iy} to higher orders of perturbation
theory, Beisert, Dippel and Staudacher proposed the following
all-loop Bethe equations \cite{Beisert:2004hm}\footnote{These
equations can be systematically derived in perturbation theory by
applying coordinate Bethe ansatz to (\ref{ham})
\cite{Staudacher:2004tk}.}:
\begin{equation}\label{asbe}
 \left(\frac{x\left(u_j+\frac{i}{2}\right)}{x\left(u_j-\frac{i}{2}\right)}\right)^L
 =\prod_{k\neq
 j}^{}\frac{u_j-u_k+i}{u_j-u_k-i}\,,
\end{equation}
where $u_j$ ($j=1,\ldots ,J_2$) are rapidities of elementary
excitations and
\begin{equation}\label{}
 x(u)=\frac{u}{2}+\frac{1}{2}\,\sqrt{u^2-\frac{\lambda }{4\pi
 ^2}}\,.
\end{equation}
The scaling dimension is given by\footnote{This is (\ref{1}) in the
rapidity parametrization: $\,{\rm e}\,^{ip}=x(u+i/2)/x(u-i/2)$.}
\begin{equation}\label{}
 \Delta =L+\frac{i\lambda }{8\pi ^2}\sum_{j=1}^{J_2}\left
 (\frac{1}{x\left(u_j+\frac{i}{2}\right)}-\frac{1}{x\left(u_j-\frac{i}{2}
 \right)}\right).
\end{equation}
The trace cyclicity of the SYM operators requires the wave function
to be periodic. This imposes an extra condition
\begin{equation}\label{}
 \prod_{j=
 1}^{J_2}\frac{x\left(u_j+\frac{i}{2}\right)}{x\left(u_j-\frac{i}{2}\right)}=1,
\end{equation}
 which makes the  total
momentum an integer multiple of $2\pi $.

These equations are asymptotic in a certain sense
\cite{Beisert:2004hm} and compute the eigenvalues of (\ref{ham}) up
to the $O(\lambda ^L)$ accuracy. The wrapping interactions
\cite{Beisert:2004hm,Beisert:2004yq}, which start to contribute at
this order of perturbation theory, may invalidate or modify the
asymptotic Bethe ansatz. These corrections, however, are
exponentially small in the thermodynamic limit of $L\rightarrow
\infty $, at least if the 't~Hooft coupling is not very large.

The energy density in the anti-ferromagnetic vacuum can be
calculated by standard techniques \cite{hul,Faddeev:1996iy}. Taking
logarithm of both sides of (\ref{asbe}) we get
\begin{equation}\label{blog}
Lp(u_j)=2\pi k_j+\sum_{k\neq j}^{}\Phi (u_j-u_k),
\end{equation}
where
\begin{equation}\label{mom}
 p(u)=\frac{1}{i}\,\ln\frac{u+\frac{i}{2}+
 \sqrt{\left(u+\frac{i}{2}\right)^2-\frac{\lambda }{4\pi ^2}}}
 {u-\frac{i}{2}+
 \sqrt{\left(u-\frac{i}{2}\right)^2-\frac{\lambda }{4\pi ^2}}}
\end{equation}
is the momentum of an elementary excitation and
\begin{equation}\label{}
 \Phi (u)=\pi -2\arctan u
\end{equation}
is the phaseshift due to pairwise scattering. It is assumed that the
same branch of the logarithm is used for all momenta in (\ref{mom}).
Let us fix the conventions by requiring that $p(u)$ and $\Phi (u)$
change from $2\pi $ at $u\rightarrow -\infty $ to $0$ at
$u\rightarrow +\infty $. The arbitrariness in choosing the branch of
the logarithm is then entirely encoded in the mode numbers $k_j$.

There is one excitation per mode number in the AF state. It then
follows from (\ref{blog}) that $k_j$ can take $L/2$ values from $1$
to $L-M=L/2$. Thus all available levels are filled and, assuming
that $u_j$ monotonously decreases with $j$, we can set $k_j=j$.
After introducing the scaling variable $\xi =j/L$ and taking the
thermodynamic limit $L\rightarrow \infty $, the Bethe equation
(\ref{blog}) can be written as
\begin{equation}\label{blog1}
 p(u(\xi ))=2\pi\xi +\int_{}^{}d\eta\,\Phi (u(\xi )-u(\eta)).
\end{equation}
Differentiating in $u $ we get
\begin{equation}\label{bfin}
 \frac{i}{2}\left[
 \frac{1}{\sqrt{\left(u+\frac{i}{2}\right)^2-\frac{\lambda }{4\pi
 ^2}}}-
 \frac{1}{\sqrt{\left(u-\frac{i}{2}\right)^2-\frac{\lambda }{4\pi ^2}}}
 \right]=\pi \rho (u)+\int_{-\infty }^{+\infty
 }\frac{dv\,\rho (v)}{(u-v)^2+1}\,,
\end{equation}
where
\begin{equation}\label{}
 \rho (u)=-\frac{d\xi }{du}\,
\end{equation}
is the density of Bethe roots.

The integral equation (\ref{bfin}) can be solved by the Fourier
transform:
\begin{equation}\label{dens}
 \rho (u)=\int_{-\infty }^{+\infty }\frac{dk}{2\pi }\,\,{\rm
 e}\,^{iku}\,\frac{J_0\left(\frac{\sqrt{\lambda }\,k}{2\pi }\right)}
 {2\cosh\frac{k}{2}}\,,
\end{equation}
where $J_0$ is the Bessel function. Plugging this solution into the
equation for the energy:
\begin{equation}\label{}
 \frac{\Delta_{\rm max} }{L}=1+\frac{i\lambda }{8\pi ^2}\int_{-\infty }^{+\infty }
 du\,\rho (u)\left
 (\frac{1}{x\left(u+\frac{i}{2}\right)}-\frac{1}{x\left(u-\frac{i}{2}
 \right)}\right),
\end{equation}
 we find:
\begin{equation}\label{ee}
 \frac{\Delta_{\rm max} }{L}=1+\frac{\sqrt{\lambda }}{\pi }\int_{0}^{\infty }
 \frac{dk}{k}\,\,\frac{J_0\left(\frac{\sqrt{\lambda }\,k}{2\pi }\right)
 J_1\left(\frac{\sqrt{\lambda }\,k}{2\pi }\right)}{\,{\rm
 e}\,^{k}+1}\,.
\end{equation}
This is the main result of this paper.

Expanding in $\lambda $ we get
\begin{equation}\label{}
 \frac{\Delta_{\rm max} }{L}=1+4\ln 2\,\frac{\lambda }{16\pi ^2}-9\zeta
 (3)\left(\frac{\lambda }{16\pi ^2}\right)^2
 +75\zeta (5)\left(\frac{\lambda }{16\pi ^2}\right)^3+\ldots
\end{equation}
The second term is the ground-state energy of the Heisenberg AF. The
third term can be computed  from (\ref{ham}) by first-order
perturbation theory, since the two-site spin correlator in the
Heisenberg model is known \cite{tak}
$$\left\langle{}_{\rm AF} 0\right|\boldsymbol{\sigma
}_l\cdot\boldsymbol{\sigma }_{l+2}\left|\,0_{\rm
AF}\right\rangle=1-16\ln 2+9\zeta (3).$$ Alternatively, the two-loop
correction can be extracted from the exact solution \cite{din} of
the Imozemtsev model \cite{Inozemtsev:2002vb} \footnote{The
three-loop dilatation operator can be also embedded into the
Inozemtsev model, but the embedding is rather non-trivial
\cite{Serban:2004jf}.}.

Extrapolating (\ref{ee}) to the strong coupling we find
\begin{equation}\label{str}
 \frac{\Delta_{\rm max} }{L}=\frac{\sqrt{\lambda }}{\pi ^2}+\ldots
\end{equation}
It is not clear how justified is this extrapolation. The string
Bethe equations \cite{Arutyunov:2004vx,Beisert:2005fw}, which are
supposed to describe the spectrum at strong coupling, contain a
correction term that definitely contributes in the thermodynamic
limit. The order of limits is also important here\footnote{I would
like to thank J.~Minahan for the discussion of this point.}. The
derivation of (\ref{str}) assumes that the limit $L\rightarrow
\infty $ is taken before $\lambda \rightarrow \infty $. The
consistency of replacing (\ref{blog}) by (\ref{blog1}) and
(\ref{bfin}) requires that the distance between nearby roots is
small: $\Delta u\sim 1/L\rho (u)\ll 1$. Since $\rho \sim
1/\sqrt{\lambda }$ at large $\lambda $, (\ref{str}) holds only for
$\lambda \ll L^2$.

It is interesting to see how the density behaves at large $\lambda
$. It will then become clear that the $\sqrt{\lambda }\,L$ scaling
of $\dm$ is the robust prediction, to a large degree independent of
a particular form of Bethe equations.
\begin{figure}[t]
\centerline{\includegraphics[width=4cm]{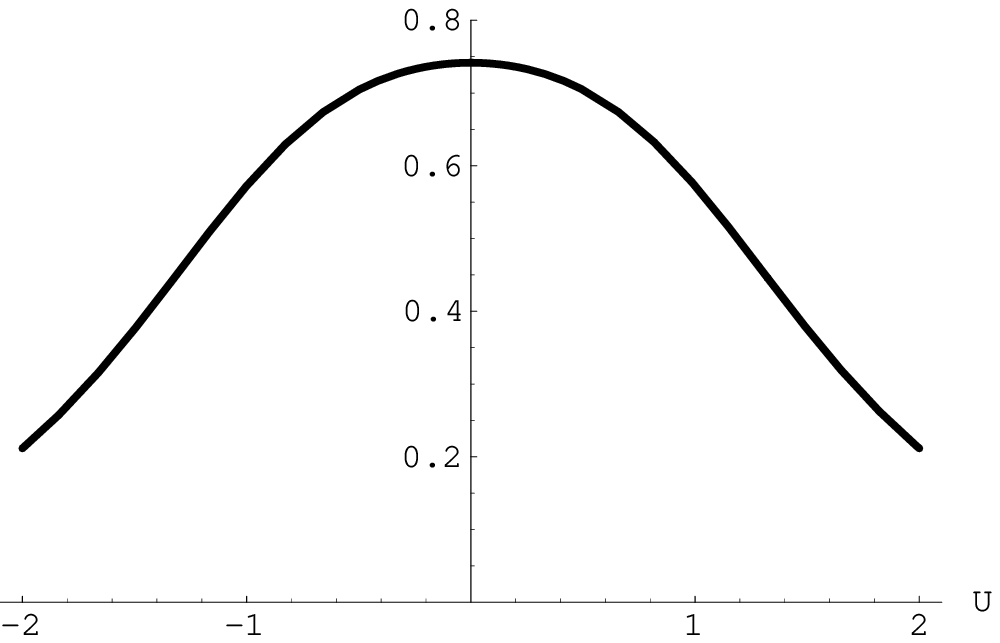}\qquad
\includegraphics[width=4cm]{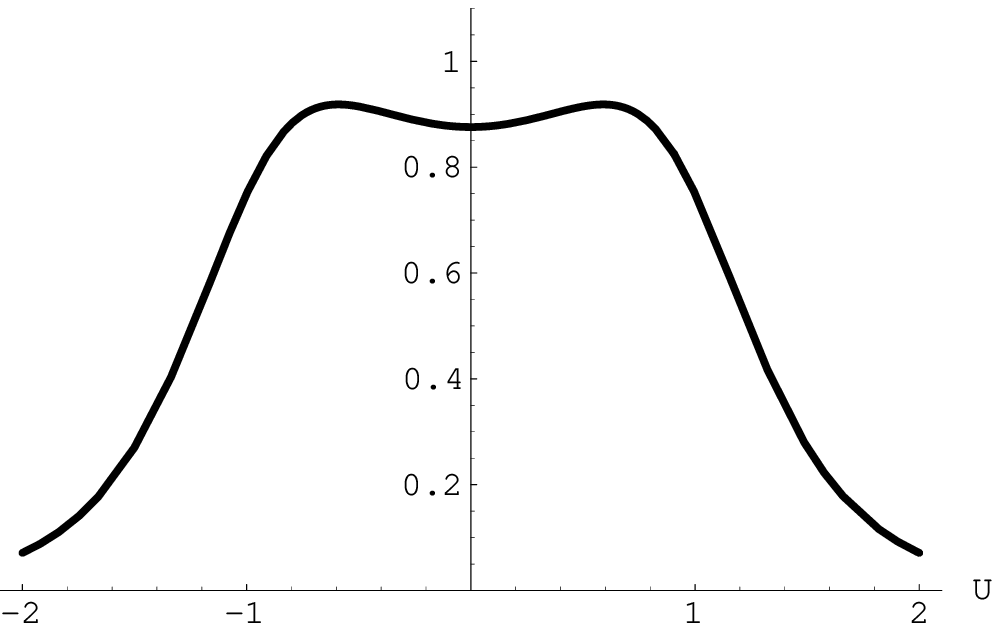}\qquad
\includegraphics[width=4cm]{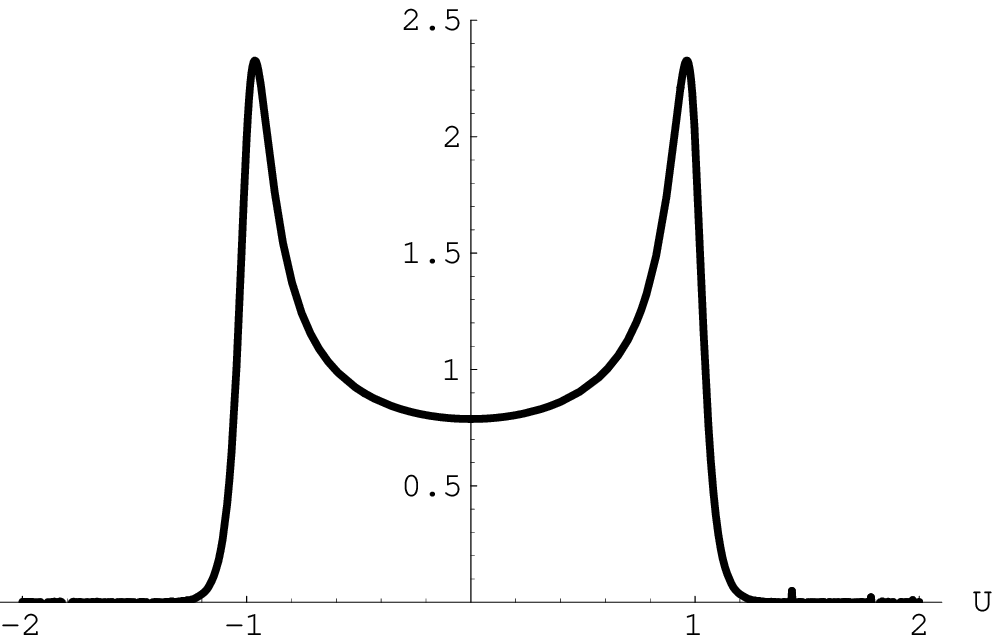}}
\caption{\label{fig1}\small The density of Bethe roots ($2\pi \rho/
\sqrt{\lambda }$) as a function of the scaling variable $U=2\pi
u/\sqrt{\lambda }$ at $\lambda =7$ (left pane); $\lambda =30$
(middle pane); and $\lambda =2000$ (right pane).}
\end{figure}
At one-loop,
\begin{equation}\label{}
 \left.\rho (u)\right|_{\lambda \rightarrow 0}
 =\frac{1}{2\cosh\pi u}\,,
\end{equation}
which monotonously decreases with $|u|$. As $\lambda $ grows the
density develops two peaks which become more and more pronounced
(fig.~\ref{fig1}). The positions of the peaks approach $u=\pm
\sqrt{\lambda }/2\pi $ at strong coupling and the density
asymptotically approaches
\begin{equation}\label{stc}
 \left.\rho (u)\right|_{\lambda \rightarrow \infty }
 =\frac{\theta \left(\frac{\lambda }{4\pi ^2}-u^2\right)}
 {2\pi \sqrt{\frac{\lambda }{4\pi ^2}-u^2}}\,.
\end{equation}
This is somewhat similar to the rapidity distribution in the
conformal supercoset $O(2m+2|2m)$ sigma-model \cite{Mann:2005ab},
where the peaks correspond to special low-energy modes coined
non-movers in \cite{Mann:2005ab}. The distribution (\ref{stc}),
however, has a very simple form in the momentum representation. At
strong coupling,
\begin{equation}\label{}
 x\left(u\pm\frac{i}{2}\right)\approx \frac{u}{2}\pm\frac{i}{2}\,
 \sqrt{\frac{\lambda }{4\pi ^2}-u^2}
\end{equation}
for $|u|<\sqrt{\lambda }/2\pi $ and
\begin{equation}\label{}
 u=\frac{\sqrt{\lambda }}{2\pi }\,\cos\frac{p}{2}\,.
\end{equation}
Changing the variables from $u$ to $p$ we find that (\ref{stc})
corresponds to the flat distribution of momenta in the whole
Brillouin zone:
\begin{equation}\label{}
 \left.\rho(p)\right|_{\lambda \rightarrow \infty }=\frac{1}{4\pi
 }\,.
\end{equation}
In other words, the Bethe equations are solved by $p_n=4\pi n/L$;
$n=1,\ldots ,L/2$. This looks like ordinary momentum quantization,
but the quantum of momentum is twice as large as allowed by the
periodic boundary conditions. This is because the scattering phase
is also non-trivial:
$$
\prod_{n=1}^{L/2}\frac{\frac{\sqrt{\lambda }}{2\pi }\left(
\cos\frac{p}{2}-\cos\frac{2\pi n}{L}\right)+i}{\frac{\sqrt{\lambda
}}{2\pi }\left( \cos\frac{p}{2}-\cos\frac{2\pi n}{L}\right)-i}
\approx -\,{\rm e}\,^{ipL/2},
$$
and forces the momentum to be quantized in the units of $4\pi /L$.

One can in principle do the same calculation for the string Bethe
ansatz of \cite{Arutyunov:2004vx}. The result should not be much
different, since the dispersion relation is the same
eq.~(\ref{1})\footnote{As a matter of fact, the string Bethe ansatz
\cite{Arutyunov:2004vx} postulates the same mechanism for the
finiteness of the number of states: the dispersion relation is
periodic in $p$, momentum is thus confined to one Brillouin zone.}.
At strong coupling it becomes
$$
\varepsilon \approx \frac{\sqrt{\lambda }}{2\pi
}\,\sin\frac{p}{2}\,.
$$
As a result, the upper bound on the energy is proportional to
$\sqrt{\lambda }$. The momentum distribution is not very important
for this conclusion. The concrete form of the distribution only
determines the coefficient of proportionality. It would be extremely
interesting to check this prediction directly from the string theory
in $AdS_5\times S^5$.

I am grateful to S.~Frolov, J.~Minahan, A.Tirziu and especially to
A.~Tseytlin for comments and discussions. I would like to thank
A.~Tirziu and A.~Tseytlin for showing me their results \cite{TT}
prior to publication. The work of K.Z. was supported in part by the
Swedish Research Council (VR) under contracts 621-2002-3920 and
621-2004-3178, by the G\"oran Gustafsson Foundation, and by RFBR
grant NSh-1999.2003.2 for the support of scientific schools.

{\bf Note added:} While this paper was being prepared for
publication, \cite{Rej:2005qt} appeared which also contains the
calculation of the energy of the AF state. The authors of
\cite{Rej:2005qt} in addition established an interesting
relationship between the spin chain (\ref{ham}) and the Hubbard
model.


\begin{thebibliography}{99}
\addtolength{\itemsep}{-6pt}


%%CITATION = HEP-TH 9711200;%%
\bibitem{Maldacena:1998re}
J.~M.~Maldacena, {"The large N limit of superconformal field
theories and
  supergravity''},
Adv.~Theor.~Math.~Phys. ~\textbf{2}~(1998)~231,
[arXiv:hep-th/9711200].

\bibitem{Gubser:1998bc}
S.~S.~Gubser, I.~R.~Klebanov and A.~M.~Polyakov, ``Gauge theory
correlators from non-critical string theory,'' Phys.\ Lett.\ B {\bf
428}, 105 (1998) [hep-th/9802109].
%%CITATION = HEP-TH 9802109;%%

%\cite{Witten:1998qj}
\bibitem{Witten:1998qj}
E.~Witten, ``Anti-de Sitter space and holography,'' Adv.\ Theor.\
Math.\ Phys.\  {\bf 2}, 253 (1998) [hep-th/9802150].
%%CITATION = HEP-TH 9802150;%%

\bibitem{Beisert:2003jj}
N.~Beisert, {``The complete one-loop dilatation operator of \4N
super Yang-Mills theory,''} Nucl.\ Phys.\ B {\bf 676}, 3 (2004)
[arXiv:hep-th/0307015].
%%CITATION = HEP-TH 0307015;%%


%\cite{Minahan:2002ve}
\bibitem{Minahan:2002ve}
J.~A.~Minahan and K.~Zarembo, {"The Bethe-ansatz for \4N super
Yang-Mills,''} JHEP {\bf 0303}, 013 (2003) [arXiv:hep-th/0212208].
%%CITATION = HEP-TH 0212208;%%

\bibitem{Beisert:2003tq}
N.~Beisert, C.~Kristjansen and M.~Staudacher, "The dilatation
operator of $\mathcal{N}=4$ super Yang-Mills theory,'' Nucl.\ Phys.\
B {\bf 664}, 131 (2003) [arXiv:hep-th/0303060].
%%CITATION = HEP-TH 0303060;%%


\bibitem{Serban:2004jf}
  D.~Serban and M.~Staudacher,
  ``Planar N = 4 gauge theory and the Inozemtsev long range spin chain,''
  JHEP {\bf 0406} (2004) 001
  [arXiv:hep-th/0401057].
  %%CITATION = HEP-TH 0401057;%%


\bibitem{Beisert:2004hm}
N.~Beisert, V.~Dippel and M.~Staudacher, ``A novel long range spin
chain and planar N = 4 super Yang-Mills,'' JHEP {\bf 0407}, 075
(2004) [arXiv:hep-th/0405001].
%%CITATION = HEP-TH 0405001;%%



%%CITATION = HEP-TH 0202021;%%
\bibitem{Berenstein:2002jq}
D.~Berenstein, J.~M.~Maldacena and H.~Nastase,
%{``Strings in flat space and pp waves from {$\mathcal{N}=\mathord{}$4}
 % {Super} {Yang Mills}''},
JHEP~\textbf{0204}~(2002)~013, [arXiv:hep-th/0202021].

\bibitem{Frolov:2003qc}
S.~Frolov and A.~A.~Tseytlin, {``Multi-spin string solutions in
{$AdS_5\times S^5$}''}, Nucl.~Phys.~\textbf{B668}~(2003)~77,
[arXiv:hep-th/0304255];
%%CITATION = HEP-TH 0304255;%%
%\bibitem{Frolov:2003xy}
%S.~Frolov and A.~A.~Tseytlin,
``Rotating string solutions: AdS/CFT duality in non-supersymmetric
sectors,'' Phys.\ Lett.\ B {\bf 570}, 96 (2003)
[arXiv:hep-th/0306143].
%%CITATION = HEP-TH 0306143;%%

\bibitem{Kruczenski:2003gt}
  M.~Kruczenski,
  ``Spin chains and string theory,''
  Phys.\ Rev.\ Lett.\  {\bf 93} (2004) 161602
  [arXiv:hep-th/0311203].
  %%CITATION = HEP-TH 0311203;%%

\bibitem{Kruczenski:2004kw}
M.~Kruczenski, A.~V.~Ryzhov and A.~A.~Tseytlin, ``Large spin limit
of {$AdS_5\times S^5$} string theory and low energy expansion of
ferromagnetic spin chains,'' Nucl.\ Phys.\ B {\bf 692}, 3 (2004)
[arXiv:hep-th/0403120].
%%CITATION = HEP-TH 0403120;%%

\bibitem{Kazakov:2004qf}
V.~A.~Kazakov, A.~Marshakov, J.~A.~Minahan and K.~Zarembo,
``Classical / quantum integrability in AdS/CFT,'' JHEP {\bf 0405},
024 (2004) [arXiv:hep-th/0402207].
%%CITATION = HEP-TH 0402207;%%

\bibitem{Minahan:2005jq}
  J.~A.~Minahan,
  ``The SU(2) sector in AdS/CFT,''
  Fortsch.\ Phys.\  {\bf 53} (2005) 828
  [arXiv:hep-th/0503143].
  %%CITATION = HEP-TH 0503143;%%

\bibitem{Gubser:2002tv}
S.~S.~Gubser, I.~R.~Klebanov and A.~M.~Polyakov, {``A semi-classical
limit of the gauge/string correspondence''},
Nucl.~Phys.~\textbf{B636}~(2002)~99, [arXiv:hep-th/0204051].
%%CITATION = HEP-TH 0204051;%%


\bibitem{Arutyunov:2004vx}
  G.~Arutyunov, S.~Frolov and M.~Staudacher,
  ``Bethe ansatz for quantum strings,''
  JHEP {\bf 0410} (2004) 016
  [arXiv:hep-th/0406256].
  %%CITATION = HEP-TH 0406256;%%

\bibitem{Beisert:2005tm}
  N.~Beisert,
  ``The su(2$|$2) dynamic S-matrix,''
  arXiv:hep-th/0511082.
  %%CITATION = HEP-TH 0511082;%%

\bibitem{Sundborg:1999ue}
  B.~Sundborg,
  ``The Hagedorn transition, deconfinement and N = 4 SYM theory,''
  Nucl.\ Phys.\ B {\bf 573} (2000) 349
  [arXiv:hep-th/9908001].
  %%CITATION = HEP-TH 9908001;%%

\bibitem{Aharony:2003sx}
  O.~Aharony, J.~Marsano, S.~Minwalla, K.~Papadodimas and M.~Van Raamsdonk,
  ``The Hagedorn / deconfinement phase transition in weakly coupled large N
  gauge theories,''
  Adv.\ Theor.\ Math.\ Phys.\  {\bf 8} (2004) 603
  [arXiv:hep-th/0310285].
  %%CITATION = HEP-TH 0310285;%%


\bibitem{Arutyunov:2005hd}
  G.~Arutyunov and S.~Frolov,
  ``Uniform light-cone gauge for strings in {$AdS_5\times S^5$}:
  Solving su(1$|$1)
  sector,''
  arXiv:hep-th/0510208.
  %%CITATION = HEP-TH 0510208;%%

\bibitem{Minahan:2005qj}
  J.~A.~Minahan, A.~Tirziu and A.~A.~Tseytlin,
  ``$1/J^2$ corrections to BMN energies from the quantum long range
  Landau-Lifshitz model,''
  arXiv:hep-th/0510080.
  %%CITATION = HEP-TH 0510080;%%

\bibitem{Gromov:2005gp}
  N.~Gromov and V.~Kazakov,
  ``Double scaling and finite size corrections in sl(2) spin chain,''
  arXiv:hep-th/0510194.
  %%CITATION = HEP-TH 0510194;%%

\bibitem{Park:2005kt}
  I.~Y.~Park, A.~Tirziu and A.~A.~Tseytlin,
  ``Semiclassical circular strings in $AdS_5$ and 'long' gauge field strength
  operators,''
  Phys.\ Rev.\ D {\bf 71} (2005) 126008
  [arXiv:hep-th/0505130].
  %%CITATION = HEP-TH 0505130;%%

\bibitem{TT}
 A.~Tirziu and A.~A.~Tseytlin, in progress.

\bibitem{Bukhvostov:1985rn}
  A.~P.~Bukhvostov, G.~V.~Frolov, L.~N.~Lipatov and E.~A.~Kuraev,
  ``Evolution Equations For Quasi - Partonic Operators,''
  Nucl.\ Phys.\ B {\bf 258} (1985) 601.
  %%CITATION = NUPHA,B258,601;%%

\bibitem{Braun:1998id}
  V.~M.~Braun, S.~E.~Derkachov and A.~N.~Manashov,
  ``Integrability of three-particle evolution equations in {QCD},''
  Phys.\ Rev.\ Lett.\  {\bf 81} (1998) 2020
  [arXiv:hep-ph/9805225].
  %%CITATION = HEP-PH 9805225;%%


\bibitem{Belitsky:1999bf}
  A.~V.~Belitsky,
  ``Renormalization of twist-three operators and integrable lattice models,''
  Nucl.\ Phys.\ B {\bf 574} (2000) 407
  [arXiv:hep-ph/9907420].
  %%CITATION = HEP-PH 9907420;%%

\bibitem{Ferretti:2004ba}
  G.~Ferretti, R.~Heise and K.~Zarembo,
  ``New integrable structures in large-N QCD,''
  Phys.\ Rev.\ D {\bf 70} (2004) 074024
  [arXiv:hep-th/0404187].
  %%CITATION = HEP-TH 0404187;%%

\bibitem{Belitsky:2004cz}
  A.~V.~Belitsky, V.~M.~Braun, A.~S.~Gorsky and G.~P.~Korchemsky,
  ``Integrability in QCD and beyond,''
  Int.\ J.\ Mod.\ Phys.\ A {\bf 19} (2004) 4715
  [arXiv:hep-th/0407232].
  %%CITATION = HEP-TH 0407232;%%

\bibitem{Beisert:2004fv}
  N.~Beisert, G.~Ferretti, R.~Heise and K.~Zarembo,
  ``One-loop QCD spin chain and its spectrum,''
  Nucl.\ Phys.\ B {\bf 717} (2005) 137
  [arXiv:hep-th/0412029].
  %%CITATION = HEP-TH 0412029;%%

\bibitem{Beisert:2003ys}
  N.~Beisert,
  ``The $su(2|3)$ dynamic spin chain'',
  Nucl.\ Phys.\ B {\bf 682} (2004) 487,
  hep-th/0310252.
  %%CITATION = HEP-TH 0310252;%%

\bibitem{Zwiebel:2005er}
  B.~I.~Zwiebel,
  ``N = 4 SYM to two loops: Compact expressions for the non-compact symmetry
  algebra of the su(1,1$|$2) sector,''
  arXiv:hep-th/0511109.
  %%CITATION = HEP-TH 0511109;%%

\bibitem{Kotikov:2004er}
A.~V.~Kotikov, L.~N.~Lipatov, A.~I.~Onishchenko and
V.~N.~Velizhanin, ``Three-loop universal anomalous dimension of the
Wilson operators in N = 4 SUSY Yang-Mills model,'' Phys.\ Lett.\ B
{\bf 595}, 521 (2004) [arXiv:hep-th/0404092].
%%CITATION = HEP-TH 0404092;%%

\bibitem{Eden:2004ua}
  B.~Eden, C.~Jarczak and E.~Sokatchev,
  ``A three-loop test of the dilatation operator in N = 4 SYM,''
  Nucl.\ Phys.\ B {\bf 712} (2005) 157
  [arXiv:hep-th/0409009].
  %%CITATION = HEP-TH 0409009;%%

\bibitem{Fischbacher:2004iu}
  T.~Fischbacher, T.~Klose and J.~Plefka,
  ``Planar plane-wave matrix theory at the four loop order: Integrability
  without BMN scaling,''
  JHEP {\bf 0502} (2005) 039
  [arXiv:hep-th/0412331].
  %%CITATION = HEP-TH 0412331;%%

\bibitem{Beisert:2005wv}
  N.~Beisert and T.~Klose,
  ``Long-range gl(n) integrable spin chains and plane-wave matrix theory,''
  arXiv:hep-th/0510124.
  %%CITATION = HEP-TH 0510124;%%

\bibitem{Bethe:1931hc}
H.~Bethe, ``On The Theory Of Metals. 1. Eigenvalues And
Eigenfunctions For The Linear Atomic Chain,'' Z.\ Phys.\  {\bf 71},
205 (1931).
%%CITATION = ZEPYA,71,205;%%

\bibitem{Faddeev:1996iy}
  L.~D.~Faddeev,
  {\it How Algebraic Bethe Ansatz works for integrable model}, in {\it
  Quantum Symmetries}, Proceedings of the Les Houches Summer School,
  Session LXIV, Les Houches, 1 August - 8 September 1995, eds:
  A. Connes, K. Gawedzki and J. Zinn-Justin,
  hep-th/9605187.
  %%CITATION = HEP-TH 9605187;%%

\bibitem{Staudacher:2004tk}
  M.~Staudacher,
  ``The factorized S-matrix of CFT/AdS,''
  JHEP {\bf 0505}, 054 (2005)
  [arXiv:hep-th/0412188].
  %%CITATION = HEP-TH 0412188;%%


\bibitem{Beisert:2004yq}
  N.~Beisert,
  ``Higher-loop integrability in N = 4 gauge theory,''
  Comptes Rendus Physique {\bf 5} (2004) 1039
  [arXiv:hep-th/0409147].
  %%CITATION = HEP-TH 0409147;%%

\bibitem{hul}
L.~Hulth\'{e}n, ``\"Uber das Austauschproblem eines Kristalles,"
Ark.~f.~Mat.~Astr.~o.~Fys. {\bf 26A:11} (1938) 1.

\bibitem{tak}
M.~Takahashi, ``Half-filled Hubbard model at low temperature,''
J.~Phys. {\bf C10} (1977) 1289.

\bibitem{din}
J.~Dittrich and V.I.~Inozemtsev, ``On the second-neighbor correlator
in 1D XXX quantum antiferromagnetic spin chain,''
arXiv:cond-mat/9706263.

\bibitem{Inozemtsev:2002vb}
  V.~I.~Inozemtsev,
  ``Integrable Heisenberg-van Vleck chains with variable range exchange,''
  Phys.\ Part.\ Nucl.\  {\bf 34} (2003) 166
  [Fiz.\ Elem.\ Chast.\ Atom.\ Yadra {\bf 34} (2003) 332]
  [arXiv:hep-th/0201001].
  %%CITATION = HEP-TH 0201001;%%

\bibitem{Beisert:2005fw}
  N.~Beisert and M.~Staudacher,
  ``Long-range PSU(2,2$|$4) Bethe ansaetze for gauge theory and strings,''
  Nucl.\ Phys.\ B {\bf 727}, 1 (2005)
  [arXiv:hep-th/0504190].
  %%CITATION = HEP-TH 0504190;%%

\bibitem{Mann:2005ab}
  N.~Mann and J.~Polchinski,
  ``Bethe ansatz for a quantum supercoset sigma model,''
  Phys.\ Rev.\ D {\bf 72} (2005) 086002
  [arXiv:hep-th/0508232].
  %%CITATION = HEP-TH 0508232;%%

\bibitem{Rej:2005qt}
  A.~Rej, D.~Serban and M.~Staudacher,
  ``Planar N=4 Gauge Theory and the Hubbard Model,''
  arXiv:hep-th/0512077.
  %%CITATION = HEP-TH 0512077;%%



\end{thebibliography}
\end{document}